\documentclass[twoside,leqno,twocolumn]{article}

\usepackage[letterpaper]{geometry}

\usepackage{siamproceedings}

\usepackage[T1]{fontenc}
\usepackage{amsfonts}
\usepackage{graphicx}
\usepackage{epstopdf}
\usepackage{subcaption}
\usepackage{enumitem}
\usepackage{algorithmic}
\ifpdf
  \DeclareGraphicsExtensions{.eps,.pdf,.png,.jpg}
\else
  \DeclareGraphicsExtensions{.eps}
\fi
\usepackage{textcomp}
\usepackage{xcolor}
\usepackage{booktabs}
\usepackage{tabularx}
\usepackage{enumitem}
\setitemize{noitemsep,topsep=0pt,parsep=0pt,partopsep=0pt}
\setenumerate{noitemsep,topsep=0pt,parsep=0pt,partopsep=0pt}

\newsiamremark{remark}{Remark}
\newsiamremark{hypothesis}{Hypothesis}
\crefname{hypothesis}{Hypothesis}{Hypotheses}
\newsiamthm{claim}{Claim}

\usepackage{amsopn}

\begin{document}

\newcommand\relatedversion{}

\title{LAPIS: A Performance Portable, High Productivity Compiler Framework}

\author{Brian Kelley\thanks{Sandia National Laboratories (\email{bmkelle@sandia.gov}, \mbox{\email{srajama@sandia.gov}}).}
    \and Sivasankaran Rajamanickam\footnotemark[1]}

\date{}

\maketitle












\begin{abstract}

Portability, performance, and productivity are three critical dimensions for evaluating a programming model or compiler infrastructure.  Several modern programming models for computational science focus on performance and portability. On the other end, several machine learning focused programming models focus on portability and productivity. A clear solution that is strong in all three dimensions has yet to emerge.
A second related problem arises when use cases from computational science converge with machine learning. The disparate popular frameworks of these fields require programmers to manually  integrate codes written in different frameworks. Finally, several programming frameworks lack easy options for extensibility as any new computer architecture change require complex changes to the programming models. We present LAPIS, an MLIR-based compiler that addresses all three of these challenges. We demonstrate that LAPIS can automatically lower sparse and dense linear algebra kernels from computational science and artificial intelligence use cases. We also show how LAPIS facilitates the integration of codes between PyTorch and Kokkos. We compare kernel performance with the default MLIR implementations on diverse architectures to demonstrate portability. By developing a dialect that is built on the principles of the Kokkos ecosystem, LAPIS also allows extensibility of the framework to new architectures.

\end{abstract}

\maketitle

\section{Introduction}

\begin{figure}
    \centering
    \includegraphics[width=0.8\linewidth]{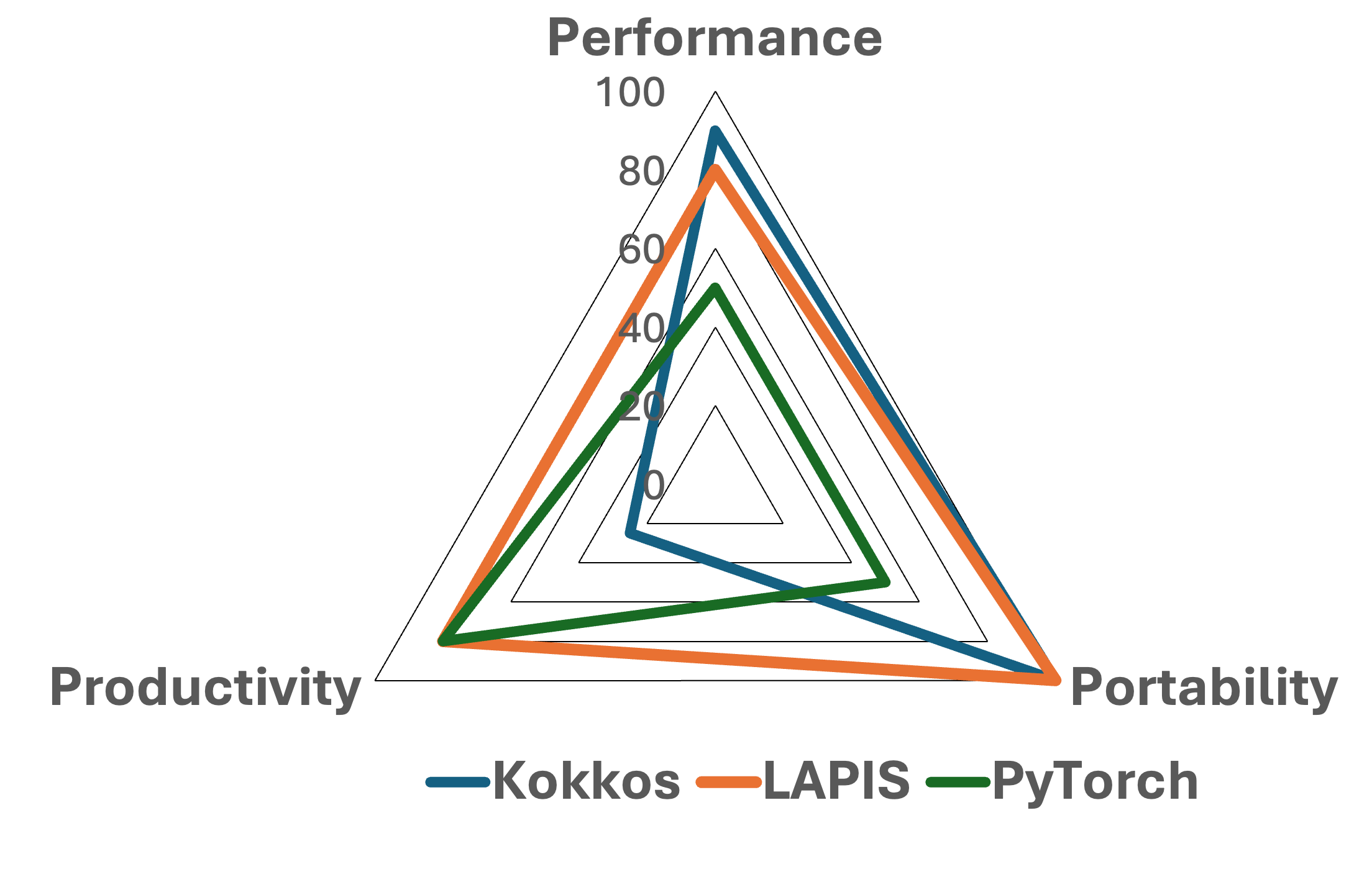}
    \caption{A notional comparison of programming frameworks for computational science and machine learning.}
    \label{fig:kokkoslapis}
\vspace{-20pt}
\end{figure}

While programming framework designs can be evaluated using many dimensions, portability, performance, and productivity remain three critical ones for long term adoption. The computational science community focusing on traditional high performance computing relies heavily on traditional programming languages such as C++ and Fortran. Programming models such as MPI, OpenMP, OpenACC or frameworks such as Kokkos \cite{kokkos}, Legion \cite{bauer2012legion}, RAJA~\cite{beckingsale2019raja}, are commonly used. Libraries such as Kokkos Kernels \cite{rajamanickam2021kokkos} and Ginkgo \cite{anzt2020ginkgo} provide a level of abstraction for sparse/dense linear algebra on a accelerators and CPUs.
Higher level frameworks such as Trilinos \cite{mayr2025trilinos} and PETSc \cite{balay2019petsc} provide one more level of abstraction for linear algebra at the distributed-memory level to the computational scientists. The most widely used frameworks like Kokkos focus on performance and portability and less on productivity (Figure \ref{fig:kokkoslapis}).

On the other end, emerging fields of artificial intelligence or machine learning (AI/ML) rely on productivity and portability focused frameworks such as PyTorch \cite{torch}, JAX or Tensorflow \cite{tensorflow}. These frameworks rely on manual interfacing to hand optimized libraries such as cuBLAS or rocBLAS for performance. The extensibility of these frameworks to new architectures is tedious limiting the architectures that are well supported by such popular frameworks. For example, emerging data flow architectures can support only a narrow sliver of these large frameworks. Such a dependence also leads to ``hardware/software lottery'' \cite{hooker2021hardware} which results in a narrow co-evolution of algorithms, libraries and hardware limiting fundamental scientific progress. The heavy reliance on dense transformer-style architectures is one recent example of such co-evolution.

We hypothesize an extensible compiler framework that can support performance, portability, and productivity will enable better computational science and rapid high performance algorithmic development within AI/ML. The development of such a compiler framework based on the lessons learned from developing Kokkos \cite{kokkos}, a performance portable programming framework for computational science and engineering (CSE), is the \textit{first challenge} addressed in this work.

Among artificial intelligence/ machine learning (AI/ML) uses, physics-informed AI/ML is becoming another tool in the tool box in the hard sciences like biology, classical mechanics, density functional theory (DFT), molecular dynamics, and chemistry. Several of these applications focus on numerical simulation of some physical phenomena. Such simulations are computationally demanding especially for high fidelity needs.

\begin{figure*}
  \centering
  \includegraphics[width=0.87\linewidth]{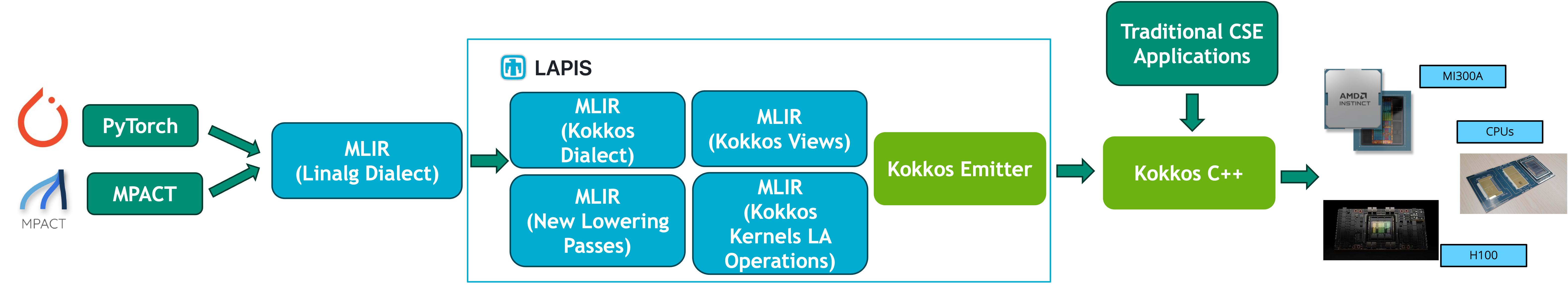}
  \caption{End-to-End LAPIS Framework from high level frameworks to different architectures. The core contributions of this work are within blue and lime green boxes.}
  \label{fig:lapis_framework}
\end{figure*}

AI/ML techniques like operator neural networks and transformers can help with some of these challenges. 
For example, ML techniques were used as a surrogate for the computational bottleneck of DFT, allowing the electronic structure of a system to be determined efficiently from just its atomic configuration \cite{ellis2021,fiedler2023predicting}.
In a multiscale simulation, the first-principle calculation is replaced by the ML surrogate within a larger-scale, lower-fidelity simulation.
Most large scale simulations are written in C++ while the ML surrogates are built using PyTorch and TensorFlow. Inter-operating between C++ and Python usually requires significant boilerplate and developer effort. The \textit{second challenge} we focus on is to interface such CSE codes with Python-based ML codes with performance close to that of native C++.
In this work, we show that a compiler can completely bypass the gap between C++ and Python-based ML codes, while yielding additional benefits in portability, efficiency and safety. Instead of explicitly calling low level functions in an ML framework from a C++ application, we allow the developer to write ML-related functions in  native Python, and then automatically compile that Python to Kokkos-based C++ source code. This can be integrated directly into a scientific application as if it had been written by hand. 

Kokkos is a shared-memory parallel programming framework for C++ that is designed for high performance across a variety of CPU and GPU architectures. To implement the compilation pipeline, we use the MLIR (Multi-Level Intermediate Representation) library within the LLVM project \cite{mlir}. We develop LAPIS, a performance portable, high productivity,  extensible, compiler framework using MLIR. LAPIS includes a new performance portability focused intermediate dialect that matches the Kokkos ecosystem, an emitter for raising the intermediate representation to Kokkos C++ code, a portable memory model that maps to multiple accelerators and several new lowering/raising passes to support performance portability. We demonstrate the benefits of LAPIS by lowering traditional CSE operations such as sparse and dense linear algebra kernels, and by demonstrating the coupling between CSE and AI/ML frameworks on SciML and traditional ML use cases. We demonstrate portable performance of such lowered code on AMD MI300A and NVIDIA H100 GPUs, and Intel Granite Rapids CPUs.
To summarize, the main contributions of this work are:
\begin{itemize}
    \item LAPIS, a performance portable, high productivity, extensible compiler framework;
    \item A performance portability focused, Kokkos ecosystem-inspired dialect, new lowering passes that correctly map hierarchical parallelism, memory model for accelerators, Kokkos Kernels linear algebra as MLIR operations, and a new emitter for parallel C++ code in the MLIR framework;
    \item Demonstration of LAPIS usage in lowering one sparse and two dense linear algebra operations on three different target architectures matching performance of hand-optimized implementations; 
    \item Demonstration of LAPIS to couple CSE and AI/ML use cases in SciML and traditional image processing use cases and demonstration of performance on three different architectures.
\end{itemize}


\section{Related Work}
\label{sec:related}

Domain specific compilers for sparse and dense operations is a rich field with several recent efforts.
The Tensor Algebra Compiler or TACO \cite{kjolstad2017tensor} is a compiler that focuses on sparse/dense tensor operations with C++ and Python interfaces. While TACO has support for generating CUDA code, it does not support AMD GPUs. LAPIS tries to avoid such extensibility issues by building on top of MLIR and by developing the Kokkos dialect so extending to new architectures is simpler. 
COMET \cite{tian2021high,mutlu2020comet} is an MLIR based compiler for sparse tensor linear algebra. COMET allows for different storage for sparse matrices with support for several front ends. COMET primarily focuses on CPU architectures. MPACT \cite{mpact} is a new compiler framework built on top of MLIR and LLVM. We support the MPACT frontend in LAPIS. Triton \cite{tillet2019triton} is an MLIR based compiler framework developed by OpenAI for ML use cases, especially highly optimized dense matrix multiplications. To our knowledge, Triton does not support sparse linear algebra. 
Triton also focuses on CUDA/HIP and hence targeted towards NVIDIA and AMD accelerators.
Data-Centric (DaCe) programming environment, uses Stateful DataFlow multiGraph (SDFG) an intermediate representation so programs can be separated from optimizations.  DISTAL \cite{yadav2022distal}, SpDistal \cite{yadav2022spdistal} are compilers that target distributed memory for dense and sparse tensors respectively. LegateSparse \cite{yadav2023legate} distributes sparse matrices  on distributed CPUs/GPUs with SciPy code as the frontend. These codes do not target performance portability and require significant development effort from the compiler developers to support new accelerators.

\section{Background}
\label{sec:background}

\paragraph{Kokkos:}
Kokkos \cite{kokkos} enables single source performance
portable codes by providing a portable programming framework for
CPUs and GPUs. It has been adopted widely in multiple CSE applications and demonstrated on petascale and exascale platforms \cite{thompson2022lammps,bettencourt2021empire,howard2018employing}. 
Kokkos has specific abstractions for execution and memory spaces. Execution spaces are selected at compile time based on the target architectures (e.g., CUDA for NVIDIA GPUs). Memory spaces represent host and accelerator memory, as well as temporary scratchpad (e.g., CUDA shared memory for use by a team of threads).
The fundamental parallel patterns in Kokkos are parallel kernels belonging to one of three patterns parallel\_for, parallel\_reduce and parallel\_scan. The fundamental data structure is a multi-dimensional arrays (\verb|Kokkos::View|). These constructs and the \verb|View| are templated on how they have to be run (Serial, OpenMP, CUDA), and where memory is located (host, device, or shared memory). Kokkos also provides the \verb|DualView|, which manages the case when data must be replicated in two different MemorySpace (Host and Device). Users are responsible for managing where the most recent updates happened and synchronizing explicitly. This performance focused option is one missing piece in the memory constructs of MLIR framework. Our custom \texttt{kokkos} dialect in LAPIS introduces the notion of DualView as a memory space for better performance of lowered code in heterogeneous systems, but without incurring significant overhead in single-memory (CPU and APU) systems.

\paragraph{Kokkos Kernels:}
Kokkos Kernels\cite{rajamanickam2021kokkos} builds on top of Kokkos to provide portable interface to sparse, dense and batched linear algebra either through custom kernels written using Kokkos or through calls to libraries such as cuBLAS or cuSparse. In either case, the user enjoys portability and performance. LAPIS also brings similar benefits to the MLIR framework by introducing common linear algebra operations in the Kokkos dialect. LAPIS can then generate calls to Kokkos Kernels from these operations if desired.

\paragraph{MLIR:} MLIR (Multi-Level Intermediate Representation) is an open-source compiler infrastructure \cite{mlir}. The focus is to make it easy to build domain-specific compilers and support multiple hardware architectures. MLIR primarily relies on \textit{intermediate representations} that are flexible concepts ranging from high level mathematical abstractions to low level abstractions for hardware architectures. MLIR is organized into \textit{dialects}, each with its own \textit{operations}, \textit{attributes} and \textit{types}. Dialects such as \texttt{tensor} or \texttt{linalg} describe high level operations. 
For example, a matrix-matrix multiplication can be expressed with a single ``operation'' from the linear algebra dialect, \texttt{linalg.matmul}.
In addition to the dialect specifications, MLIR includes built-in compiler transformations to replace operations with equivalent but lower-level code (\textit{lowering}). For example, a \verb|linalg.matmul|  might be replaced with a triply-nested for loop that computes  scalar matrix multiplication.

Frameworks like PyTorch \cite{torch} and TensorFlow \cite{tensorflow} are early users of MLIR.  Torch-MLIR \cite{torchmlir}, an extension for PyTorch provides the functionality to generate an MLIR function from an ML model.
Many built-in MLIR dialects such as \texttt{x86vector} and \texttt{amdgpu} are focused on specific architectures differing also in functionality and performance. Other dialects can represent programming models like OpenMP and OpenACC, but MLIR lacks passes that can finish lowering from these dialects to an executable or source code. The fundamental gap addressed by this work is a pipeline that is \textit{automatic, end-to-end, and performance-portable}.

\section{LAPIS Framework}
\label{sec:lapis}

The LAPIS framework provides two main features. First, it provides an automatic lowering pipeline from high-level MLIR to portable, parallel Kokkos C++ source code. Second, it provides a custom Kokkos dialect with operations that allow Kokkos constructs and Kokkos Kernels functionality as linear algebra ``operations'' to be represented directly in MLIR. We describe the high-level design before we dive into the details of our five main changes to MLIR.

\paragraph{LAPIS Inputs:} Generally, the high-level MLIR that LAPIS accepts as input should conform to the \textit{linalg-on-tensors} set of dialects. This is one of the main ``backend contracts'' supported by torch-mlir \cite{torchmlir}, but it is broad enough to describe programs beyond just ML models. It represents computations and tensors using the \texttt{linalg} and \texttt{tensor} dialects, respectively. Attributes from the \texttt{sparse\_tensor} dialect can also be used with tensor types to specify various sparse formats. 
For the purposes of this work, LAPIS supports two frontends that both conform to linalg-on-tensors - torch-mlir and MPACT. For PyTorch-based programs with dense tensors, we use PyTorch's model exporter torchscript combined with the torch-mlir package. For both dense and sparse programs written as Torch modules, we can also use the MPACT compiler \cite{mpact}. MPACT uses the TorchFX model exporter but also does some initial lowering to reach the linalg-on-tensors level.

\begin{table*}[h]
  \caption{Examples of common MLIR operations and their possible equivalents in Kokkos C++, after lowering and emitting. The first three are built-in and the rest are part of our Kokkos dialect. As in LLVM, names preceded by \% in MLIR denote SSA (static single assignment) values. Type annotations are omitted in the MLIR shown, but are always present in real code.}
  \label{tab:mlir-kokkos-mapping}
  \centering
  \begin{tabular}{ll}
    \toprule
    MLIR     &  Kokkos C++ \\
    \midrule
    \verb|memref.store %100 %A[%1]| & \verb|A(i) = 100;| \\
    \verb|%0 = memref.alloc() : memref<200xf32>| & \verb|Kokkos::View<float[200]> v(...);| \\
    \verb|%a = math.sqrt %b| & \verb|float a = Kokkos::sqrt(b);| \\
    \verb|kokkos.range_parallel(%arg1) -> (%1) ...| & \verb|parallel_for(RangePolicy<>(N) ...| \\
    \verb|kokkos.sync %v {Device}| & \verb|v.syncDevice();| \\
    \verb|kokkos.gemm(%A, %B, %C)| & \verb|KokkosBlas::gemm(A, B, C)| \\
    \verb|kokkos.gemv(%A, %x, %y)| & \verb|KokkosBlas::gemv(A, x, y);| \\
    \bottomrule
  \end{tabular}
\end{table*}

\begin{table*}[h]
  \caption{A list of the most significant lowering transformations \textbf{developed in LAPIS} to lower from the \texttt{linalg-on-tensors} frontend dialects to the Kokkos dialect. This does not include the built-in MLIR passes used in the LAPIS pipeline.}
  \label{tab:lowering-passes}
  \centering
  \begin{tabularx}{\textwidth}{lX}
    \toprule
    Transformation Name & Effect \\
    \midrule
    \verb|linalg-to-kokkoskernels| & Replace specific linear algebra operations with Kokkos Kernels calls, e.g.\texttt{linalg.matmul} becomes \texttt{kokkos.gemm} \\
    \verb|dense-linalg-to-parallel-loops| & Replace linear algebra operations with nested \texttt{scf.parallel} loops and Kokkos-supported reductions. \\
    \verb|kokkos-loop-mapping| & Replace \texttt{scf.parallel} loop nests with Kokkos dialect parallel loops, insert team synchronization, and generate performance heuristics for Kokkos TeamPolicy parameters. \\
    \verb|kokkos-dualview-management| & Insert \texttt{kokkos.sync} and \texttt{kokkos.modify} operations for DualView-typed arrays (memrefs) according to how they are accessed and modified. \\
    \bottomrule
  \end{tabularx}
\end{table*}


\paragraph{LAPIS Dialects and lowering:}
To establish an equivalence between Kokkos constructs, Kokkos Kernels operations and MLIR operations, we were motivated to design a custom Kokkos dialect (described in detail below). Table \ref{tab:mlir-kokkos-mapping} shows the equivalence between some MLIR operations and Kokkos/Kokkos Kernels features.
Then we developed a sequence of passes to lower built-in dialects to the Kokkos dialect.
Table \ref{tab:lowering-passes} describes the effects of some of these passes. We do not describe the details of each one of these lowering passes here for conciseness but describe few other lowering passes below. Once provided with an input program, LAPIS applies a pipeline that lowers results in IR using mid-level dialects that are platform-agnostic. These include \texttt{memref} and \texttt{scf}, but also our custom \texttt{kokkos} dialect. These dialects can then be mapped to lower level architecture specific calls (as we show in our linear algebra kernels), higher level Kokkos-based C++ (as we show in our AI use cases) or lowered to architecture-specific executable if desired (as we show in our sequential x-86 use cases).

LAPIS adds five significant contibutions to MLIR: (i) the development of a Kokkos dialect (ii) mapping of Kokkos Kernels linear algebra operations to sparse and dense tensors (iii) lowering passes to the Kokkos dialect to support accelerators and (iv) a Kokkos inspired memory reference model are part of core LAPIS functionality. The fifth is a Kokkos emitter that converts this mid-level Kokkos IR directly to C++ source code. While the first four contributions to MLIR help portability, this last one is critical to our target use case of coupling AI/ML use cases with C++ simulation frameworks, as the raised C++ code can be easily integrated and maintained with the simulation codes. 

Figure \ref{fig:lapis_framework} shows the end-to-end LAPIS pipeline. We describe our five main contributions below.

\subsection{Kokkos Dialect}

LAPIS includes a custom \texttt{kokkos} dialect to represent common Kokkos constructs directly in MLIR. Like Kokkos itself, this dialect uses portable abstractions for execution and memory. In Kokkos, the basic abstraction for executable kernels is the \texttt{parallel\_for} which executes over a \textit{policy}. The policy can describe a multidimensional rectangular iteration space. A parallel loop can also perform one or more reductions using an associative join operator, making it a \texttt{parallel\_reduce}. Top-level policies are templated on \textit{execution space}, which describes where the kernel should run (i.e., GPU or CPU). These parallel loops can also nest up to three levels deep. On GPUs, Kokkos maps this maximum level of nesting to grid, block, and thread dimensions (using CUDA terminology). On CPUs, the first two dimensions map to threads and the innermost dimension maps to vectorizable for-loops. In Kokkos, the policies corresponding to the three nesting levels are called \texttt{TeamPolicy}, \texttt{TeamThreadRange} and \texttt{ThreadVectorRange}. Parallel loops without nesting can use the simpler \texttt{RangePolicy} for 1D iteration spaces and \texttt{MDRangePolicy} for higher-dimensional spaces. The \texttt{kokkos} dialect in LAPIS provides three operations for parallel loops, as well as attributes to specify nesting levels. Together, these cover all possible policy and loop nesting structures allowed by Kokkos.

\begin{itemize}
    \item \texttt{range\_parallel}: a \texttt{RangePolicy} or a \\ \texttt{MDRangePolicy}. Accepts \texttt{executionSpace} attribute and \texttt{parallelLevel} attribute to specify nesting level.
    \item \texttt{team\_parallel}: a \texttt{TeamPolicy} loop, for the outermost loop of a nested kernel. Accepts optional \textit{team size} and \textit{vector length} hints as operands, which allow these policy parameters to be computed at runtime based on heuristics.
    \item \texttt{thread\_parallel}: This represents a recurring pattern in Kokkos for two-level parallelism with a \texttt{TeamPolicy} and a \texttt{TeamThreadRange} that iterates over an iteration space together and allow the inner loop to use \texttt{ThreadVectorRange}. This also accepts a \textit{vector length} hint operand.
\end{itemize}

Furthermore,  the following operations are defined in the \texttt{kokkos} dialect for synchronization:
\begin{itemize}
    \item \texttt{single}: execute a block only once within a team, or within a thread. The level is controlled with an attribute.
    \item \texttt{team\_barrier}: synchronize all the threads within a team.
\end{itemize}

Finally, we provide operations to represent calls to the Kokkos Kernels library, which exposes portable interfaces to high-performance vendor libraries for linear algebra and other algorithms:

\begin{itemize}
    \item \texttt{gemm}: Dense matrix-matrix multiplication.
    \item \texttt{gemv}: Dense matrix-vector multiplication.
\end{itemize}

Altogether, these operations allow for lowering high-level tensor operations for sparse/dense linear algebra and AI/ML kernels to performance portable code.

\subsection{Lowering to Kokkos Dialect}
Lowering from the built-in \texttt{scf} dialect to \texttt{kokkos} takes place over several passes. These are the final three passes in the overall pipeline, and result in IR that is compatible with the Kokkos C++/emitter. Here we give particular attention to the \texttt{kokkos-loop-mapping} pass. It is responsible for choosing how to map \texttt{scf.parallel} loops to Kokkos, computing heuristic values for \texttt{Kokkos::TeamPolicy} parameters (team size and vector length), and automatically inserting team-level synchronization operations. These heuristics are informed by our own experience writing and tuning Kokkos implementations of various tensor and graph algorithms for CPUs and GPUs.

\paragraph{Loop translation:}
In the first step, the pass decides whether a given parallel loops can be executed on a GPU at all. Kokkos does not allow explicit memory management (e.g., \texttt{memref.alloc}) in device code, so loops that use it must execute on the host/CPU. Next, the rewriting of \texttt{scf.parallel} loops to Kokkos loops is decided based on the loop's maximum nesting depth:
\begin{itemize}
    \item 1: Use \texttt{range\_parallel}.
    \item 2: Use \texttt{thread\_parallel} / ThreadVector-level \texttt{range\_parallel} for the outer / inner loop. The outer loop uses both TeamPolicy and TeamThread loops to iterate.
    \item 3 or more: Use \texttt{team\_parallel} for the outer loop, a TeamThread-level \texttt{range\_parallel} for the second loop, and a ThreadVector-level \texttt{range\_parallel} for the innermost loop. All loops between the second and last are converted to sequential loops.
\end{itemize}
This cases are based on the fact that Kokkos hierarchical parallelism supports at most three levels. In the case with nesting level of at least 3, we always make the innermost (ThreadVector) loop parallel to improve memory coalescing on GPUs. In hardware, this level corresponds to GPU warps or sub-warps, depending on the vector length described below.

\paragraph{Parallelism estimation:}
In the second step, we analyze how a loop's iteration bounds are computed and try to estimate how much parallelism it uses. This is crucial for generating performant Kokkos, especially for the vector length argument of \texttt{team\_parallel} and \texttt{thread\_parallel}. If the vector length is too small, GPU code that could benefit from memory coalescing will use only a portion of each cache line. But if the vector length is too large for the workload, then some threads in each warp will sit idle. In MLIR, iteration bounds may be compile-time constants, which makes this estimation straightforward. But in other common cases, the bounds may depend on the dimensions of dynamically-shaped \texttt{memref}s. In this case, we can detect common patterns and insert code to compute this estimate at runtime. We demonstrate this with an example. An especially beneficial pattern is the CSR-like iteration pattern, shown in pseudocode below. Recall that the previous pass has rewritten all parallel loops to start at 0.  In the case of the sparse \texttt{linalg.matvec} operation with the CSR format, given the number of rows (\verb|%N|), the current row (\verb|%i|), the number of entries in the current row (\verb|%len|), and the index at which to read the matrix value and column index (\verb|%index|),
the basic code structure that is generated for compressed-sparse row (CSR) or compressed-sparse column (CSC) parallel traversal is as follows. 
\\

\vspace{-10pt}
\begin{verbatim}
scf.parallel %i from #0 to %N {
  %i_plus_one = arith.add %i, #1
  %begin = memref.load %offsets[%i]
  %end = memref.load %offsets[%i_plus_one]
  %len = arith.sub %end, %begin
  scf.parallel %j from #0 to %len {
    %index = arith.add %begin, %j
    %val = memref.load %values[%index]
    ...
  }
}
\end{verbatim}
\vspace{-5pt}

In this case, our best estimate for the parallelism of the inner loop (which will become a ThreadVector \texttt{range\_parallel}) is the average number of entries per row, rounded up. Let $k=$ \verb|%offsets[%N]|. Then this average is simply $\lceil k/N\rceil$, or equivalently $(k+N-1)/N$. This value can be arbitrarily large while the actual vector length is limited by hardware in most backends (for example, the warp size 32 on CUDA), so we later insert C++ code to clamp the vector length to the maximum for the enabled backend. If no hint for team size and/or vector length could be found, then the default value of 0 is used in the IR. In the generated C++ code, this becomes the Kokkos-defined default.

\paragraph{Synchronizations:}
The last step of \texttt{kokkos-loop-mapping} is to insert synchronization operations. \texttt{Kokkos::single()} is used in Kokkos to ensure that code with side effects is executed exactly once within a team or thread. After deciding how to map the \texttt{scf.parallel} loops, we look for operations with side effects that are inside some parallel loop but not inside the innermost parallel loop. These operations are moved inside a new \texttt{single} operation. We also conservatively generate \texttt{team\_barrier} operations inside \texttt{team\_parallel} bodies, after each TeamThread-level \texttt{range\_parallel} operation that does not perform a reduction. (as reductions imply synchronization already). This is not necessary for ThreadVector-level loops since they have implicit (SIMD-like) synchronization on all Kokkos backends.

\subsection{Kokkos inspired memory references}
\label{sec:mem}
MLIR's \texttt{memref} dialect supports an optional memory space attribute as part of the \texttt{memref} type. When lowering to a heterogeneous target system, this can explicitly mark a buffer as living in host or device memory. However, MLIR has only limited built-in capabilities for automatically managing the migration of memory between spaces, either at compile-time or runtime. The \texttt{sparse-gpu-codegen} pass lists the buffers accessed by a GPU kernel, and copies \textbf{all of them} from host to device before launching the kernel. Then it copies \textbf{all} possibly-modified buffers back to host after the kernel finishes. This generates redundant host-device transfers for programs with multiple GPU kernels without any intervening usage on host significantly affecting performance.

\paragraph{LAPIS DualViews:}
LAPIS solves the above problem with \texttt{LAPIS::DualView}, a lightweight runtime data structure to efficiently manage a buffer that may be used on both host and device. Like the existing \texttt{Kokkos::} \texttt{DualView}, on heterogeneous backends \texttt{LAPIS::DualView} contains a host buffer and device buffer with the same data type and dimensions. On backends with only one memory space, the host and device buffers are the same. \texttt{DualView} also contains flags for whether each buffer has modifications that are not reflected in the opposite buffer.

Like \texttt{sparse-gpu-codegen}, the final pass in the LAPIS pipeline \texttt{kokkos-dualview-management} scans the program for where each \texttt{memref} is accessed. Along with \texttt{host} and \texttt{device}, \texttt{DualView} is a memory space attribute defined in the Kokkos dialect. Each \texttt{memref} is assigned one of these spaces. Instead of generating eager data copies as in baseline MLIR, our pass inserts lazy copies (\texttt{kokkos.sync}) that only occur if the source buffer has actually been modified since the previous sync. And instead of eagerly copying modified buffers back to host after a GPU kernel (or host access), we insert the corresponding \texttt{kokkos.modify} operation to set the modified flag. When no data transfer is actually necessary, the overhead of a \texttt{sync} operation is that of checking a boolean flag.

\paragraph{LAPIS memory management:}
\texttt{LAPIS::DualView} also handles memory management and \texttt{memref} dialect subviewing and casting operations. The subview or cast buffer (``child'') must alias the original buffer (``parent'') for correct semantics, so \texttt{LAPIS::DualView} stores the tree of parent-child relationships. To ensure that multiple children with the same parent maintain consistent contents, children share modified flags with their parents. Calling \texttt{sync} on a child \texttt{DualView} will call \texttt{sync} on its parent. \texttt{LAPIS::DualView} also uses reference-counting with \texttt{std::shared\_ptr} to make sure that the underlying memory allocation (owned by some parent) stays in scope as long as any of its children are alive.

\subsection{Kokkos Emitter}

After lowering a program to the Kokkos dialect, the last step in the end-to-end pipeline is to convert the MLIR to C++. MLIR includes a C++ emitter for non-parallel code, and we used this as a starting point for our parallel, performance portable,  Kokkos emitter. Like the built-in C++ emitter, our Kokkos emitter performs an in-order walk of the MLIR syntax tree and emits the C++ code for one operation at a time. The SSA form of MLIR makes this straightforward – we store the result of each operation in a new C++ variable, and later rely on the C++ compiler’s optimization to keep variables alive for only the duration they are needed. The one exception is that for scalar constants (\verb|arith::constant|), each reference to the SSA variable is replaced by its value as a literal. This improves performance on GPU backends because the compiler does not propagate host constants into device code. Assigning one \verb|Kokkos::View| to another, or one \verb|LAPIS::DualView| to another, does an inexpensive shallow copy operation with reference counting, so memory safety is maintained.

There are few small necessary additions to ensure this code functions correctly as a standalone C++ code.
Kokkos needs to be explicitly initialized before using its functionality, and finalized before a program exits. Globally scoped Views (such as the weights of a model) also need to be allocated and populated before running inference, and Views in GPU memory cannot be allocated until after Kokkos has been initialized. The Kokkos emitter generates additional functions \texttt{lapis\_initialize()} and \texttt{lapis\_finalize()} to perform these steps, and it is the user's responsibility to call them in their own C++ program.
The overall LAPIS pipeline is complete once the Kokkos C++ code has been written to a file.

\section{Integration of AI/ML models with CSE codes}
\label{sec:aiml_demo}

The goal of this sub-section is to investigate whether LAPIS could enable Python-C++ interoperability, especially for use cases such as in multiscale simulations utilizing surrogate models as sub-components. This allows us to focus primarily on inference. For example, a surrogate for a high-fidelity simulation can be trained offline
and the trained model can be deployed within a lower-fidelity simulation framework.

To show that LAPIS approach is actually feasible, we set out to construct a working compilation pipeline that can automatically expose some non-trivial machine learning functionality to a C++ application. We use the ResNet18 architecture as an example to do this demonstration. ResNet-18 \cite{he2016deep} is a deep convolutional neural network (CNN) architecture. The ResNet family of models 
use skip connections to train deep convolutional neural networks for image processing tasks. We use the lowering of ResNet18 as an exemplar for the use case when images from scientific experiments are analyzed using traditional ML approaches within a larger code base integrating experimental analysis with other methods.
While we show the conventional ML approach as an exemplar here,
the same steps work for the production scientific machine learning use case for electronic structure calculations in the MALA framework \cite{cangi2024materials} (demonstrated below).

Creating an MLIR module from Python code (or code in any other programming language) is the responsibility of an external front end. 
\texttt{torch-mlir} provides a full example where the pre-trained ResNet18 image classification model is converted from PyTorch to freestanding MLIR, progressively lowered through a pipeline of built-in MLIR transformations, and finally just-in-time (JIT) compiled to native serial code using LLVM. Here, ``freestanding'' means that the MLIR code has no external dependencies - it no longer involves Python and it includes all constant data needed by the model such as the CNN weight matrices. \textit{Our objective was  to replicate this example, but generating parallel, Kokkos-based C++ code that is also performance portable}.

For the purposes of testing the generated C++ code from the Kokkos emitter, a Kokkos backend class was created in Python. This backend works as a drop-in replacement for the JIT RefBackend of torch-mlir. Given an MLIR module from the PyTorch frontend and a location where Kokkos is installed, our backend automatically goes through the following steps needed to actually run the generated code:
\begin{enumerate}
\item Execute the lowering pipeline described above
\item Emit Kokkos C++ to a file in a known location
\item Compile the Kokkos C++ into a shared library with the addition of a CTypes-friendly wrapper while maintaining tensors in the form of raw host pointers and sizes, and without mangling function names.
\item Generate a Python wrapper module using the CTypes API, which accepts tensors as NumPy arrays
\item Import the module and loading the shared library and call \texttt{lapis\_initialize()}.
\end{enumerate}

This extended pipeline is useful for testing because the resulting wrapper module provides an identical interface to the example PyTorch to the RefBackend (LLVM-based JIT) backend. In the case of the ResNet18 example, using our code is as simple as:

\begin{verbatim}
import kokkosModule
probabilities = kokkosModule.forward(image)
\end{verbatim}
where ``image'' is a preprocessed NumPy array representing the pixel values of the input, and ``probabilities'' is the resulting vector of probabilities that the image belongs to each class. This vector can be directly compared with the one returned by the interpreted PyTorch model, or the LLVM JIT version of the model.

For integration of a model into an scientific application, it is desirable to generate code with a dynamic batch size so that the same code can be called with varying problem sizes. In the case of ResNet18, the resulting function signature will be
\begin{verbatim}
LAPIS::DualView<float**, Kokkos::LayoutRight>
forward(LAPIS::DualView<float****,
  Kokkos::LayoutRight> v1);
\end{verbatim}
where \texttt{v1} contains the input images, and for a batch of N images will have dimension $N \times 3 \times 224 \times 224$. Unlike Torch, Kokkos does not allow mixed static and dynamic dimensions in any order, so LAPIS uses all-dynamic dimensions (represented by \verb|*|) for these types. \texttt{LAPIS::DualView} has a constructor that accepts either a host or device View, so this function is effectively compatible with both types. In an application, the full usage of the generated code might look like:
\begin{verbatim}
// Include the LAPIS-generated file
#include "resnet18.hpp"
...
Kokkos::View<float****, Kokkos::LayoutRight>
  images("images", N, 3, 224, 224);
auto result = forward(images);
result.syncDevice();
auto result_dev = result.device_view();
// use result_dev
\end{verbatim}

LAPIS generates the C++ file (e.g. \texttt{resnet18.hpp}) with no dependencies besides Kokkos. All model weights are included in the file as constant arrays. This eliminates the need to maintain specific versions of Python or PyTorch for calling the model within the C++ application.
The workflow is exactly the same for the DNN surrogate for electronic structure calculations in MALA code and integrating the surrogate with molecular dynamics simulations in LAMMPS \cite{thompson2022lammps}.

\section{Experimental Results}
\label{sec:results}

\subsection{Experimental setup}

To demonstrate the performance-portability of the code generated by LAPIS, we selected three modern systems that are representative of AMD GPUs used in El Capitan, the fastest supercomputer, NVIDIA GPUs the most commonly used accelerator in the Top500 and CPUs still common in many production environments. These also represent three of the most commonly used Kokkos backends: OpenMP, CUDA and HIP. 
More information on the systems, software and experimental setup are provided in \ref{sec:sysconfig}.

\subsection{Sparse linear algebra}
\label{sec:sparse}

\begin{table}[]
\begin{tabular}{llll}
Matrix           & Rows & $\mathit{nnz}_\mathit{mean}$ & $\mathit{nnz}_\mathit{max}$ \\ \hline
StocF-1465       & 1465137      & 14.34  & 189                                                           \\
PFlow\_742       & 742793       & 50     & 137                                                           \\
Elasticity3D & 648000       & 78.33  & 81                                                            \\
audikw\_1        & 943695       & 82.28  & 345                                        

\end{tabular}
\caption{Matrices used to evaluate SpMV performance in Figure \ref{fig:spmv_perf}. $\mathit{nnz}_\mathit{mean}$ and $\mathit{nnz}_\mathit{max}$ are the mean and maximum number of nonzeros per row, respectively.}
\label{tab:matrices}
\end{table}

We evaluate the LAPIS sparse tensor lowering pipeline using a fundamental linear algebra kernel - Sparse Matrix Vector multiplication (SpMV) to evaluate the portability and performance of the LAPIS generated code.  SpMV computes $y := Ax$ where $x$ and $y$ are dense vectors. We assume $A$ is a CSR-formatted sparse matrix as that is the most widely used format across high performance scientific codes. The code generated for SpMV is from a simple \texttt{torch.mv(A, x)} where \texttt{A} is of type \texttt{torch.sparse\_csr\_tensor}  and \texttt{x} is a \texttt{torch.tensor}.

LAPIS generates a fully portable code across three different architectures that achieves performance close to the achievable bandwidth limit based on comparison to the Kokkos Kernels library \cite{rajamanickam2021kokkos}. Note that this is made possible by using the hierarchical parallelism within LAPIS's Kokkos dialect, as well as the team size and vector length heuristics that imitate expert-written Kokkos code.

Table \ref{tab:matrices} lists the four matrices that were used to evaluate the performance of the LAPIS-generated SpMV kernel. All matrices except for Elasticity3D\_60 were obtained from the SuiteSparse Matrix Collection \cite{suitesparse}. Elasticity3D\_60 was generated by the Galeri package of Trilinos \cite{mayr2025trilinos} to have a $60^3$ regular grid and 3 degrees of freedom per grid point.

\begin{figure}[h]
    \centering
    \includegraphics[width=0.9\linewidth]{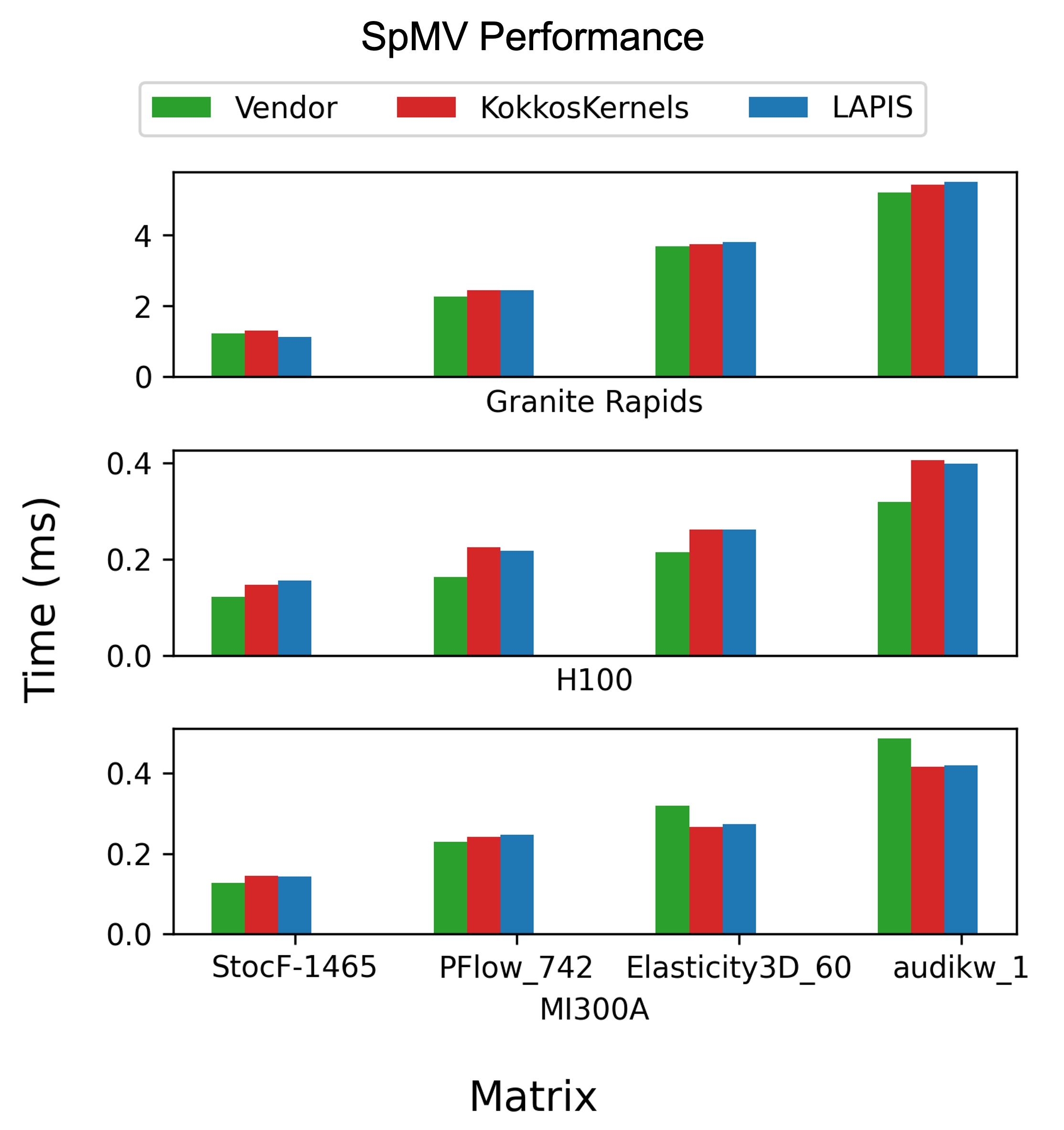}
    \caption{SpMV performance (lower is better) across three architectures and four matrices.}
    \label{fig:spmv_perf}
\end{figure}

Figure \ref{fig:spmv_perf} demonstrates the performance and portability of the SpMV kernel generated by LAPIS from the single line python call for the four test matrices. The performance of the LAPIS-generated SpMV is similar to the handwritten Kokkos-based implementation in Kokkos Kernels \cite{rajamanickam2021kokkos}. While LAPIS is somewhat slower than the cuSPARSE SpMV, it performs similarly to MKL (on Granite Rapids) and rocSPARSE (on the MI300A).

\subsection{CSE and AI/ML integration experiments}

This section shows LAPIS usage in a Scientific machine learning (SciML) use case (MALA) and a traditional machine learning use case (image processing using ResNet18).

\begin{figure}
    \centering
    \begin{subfigure}[b]{0.24\textwidth} \includegraphics[width=\textwidth]{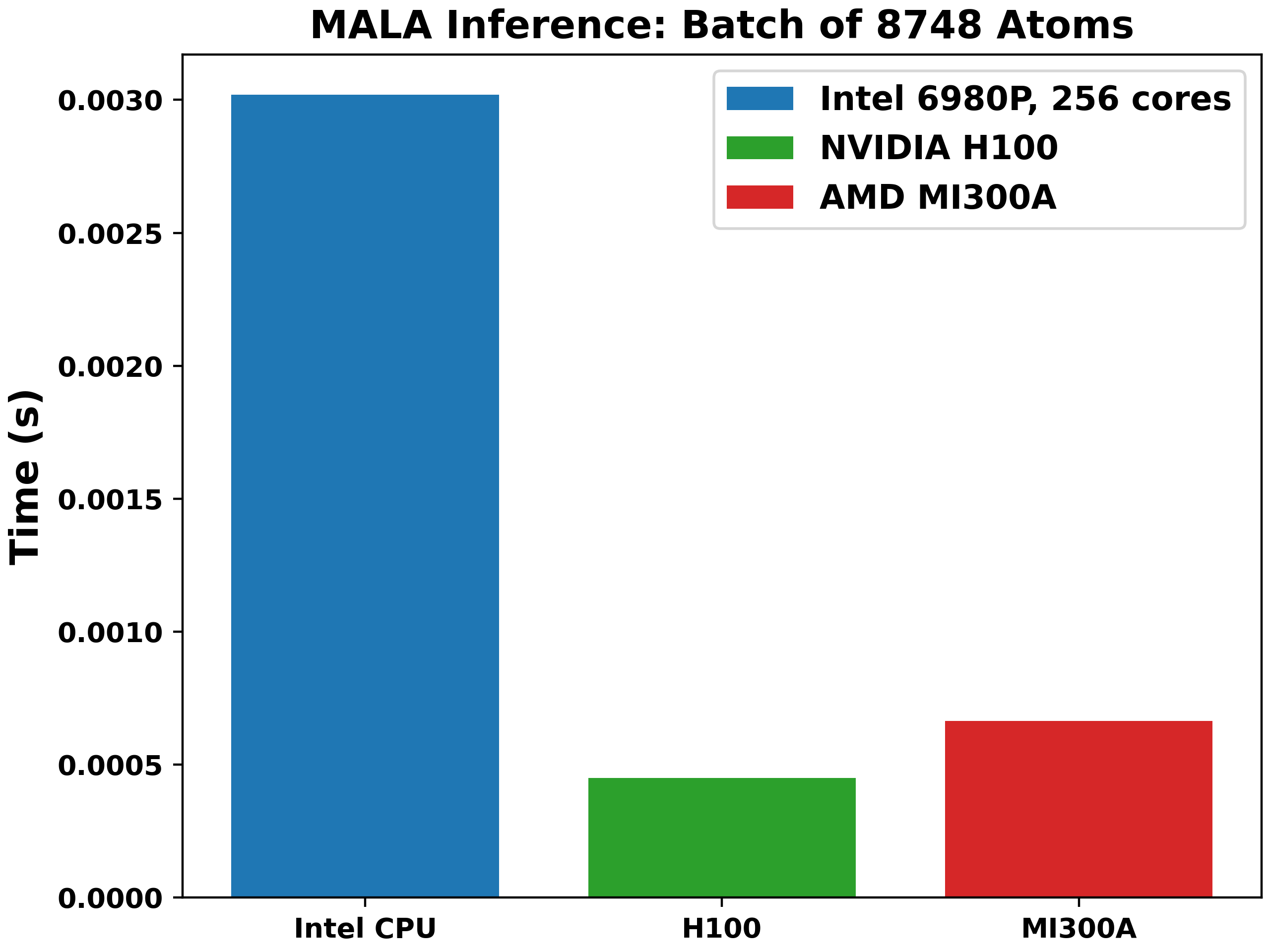}
    \caption{MALA inference}
    \label{fig:mala}    
    \end{subfigure}
    \begin{subfigure}[b]{0.24\textwidth} \includegraphics[width=\textwidth]{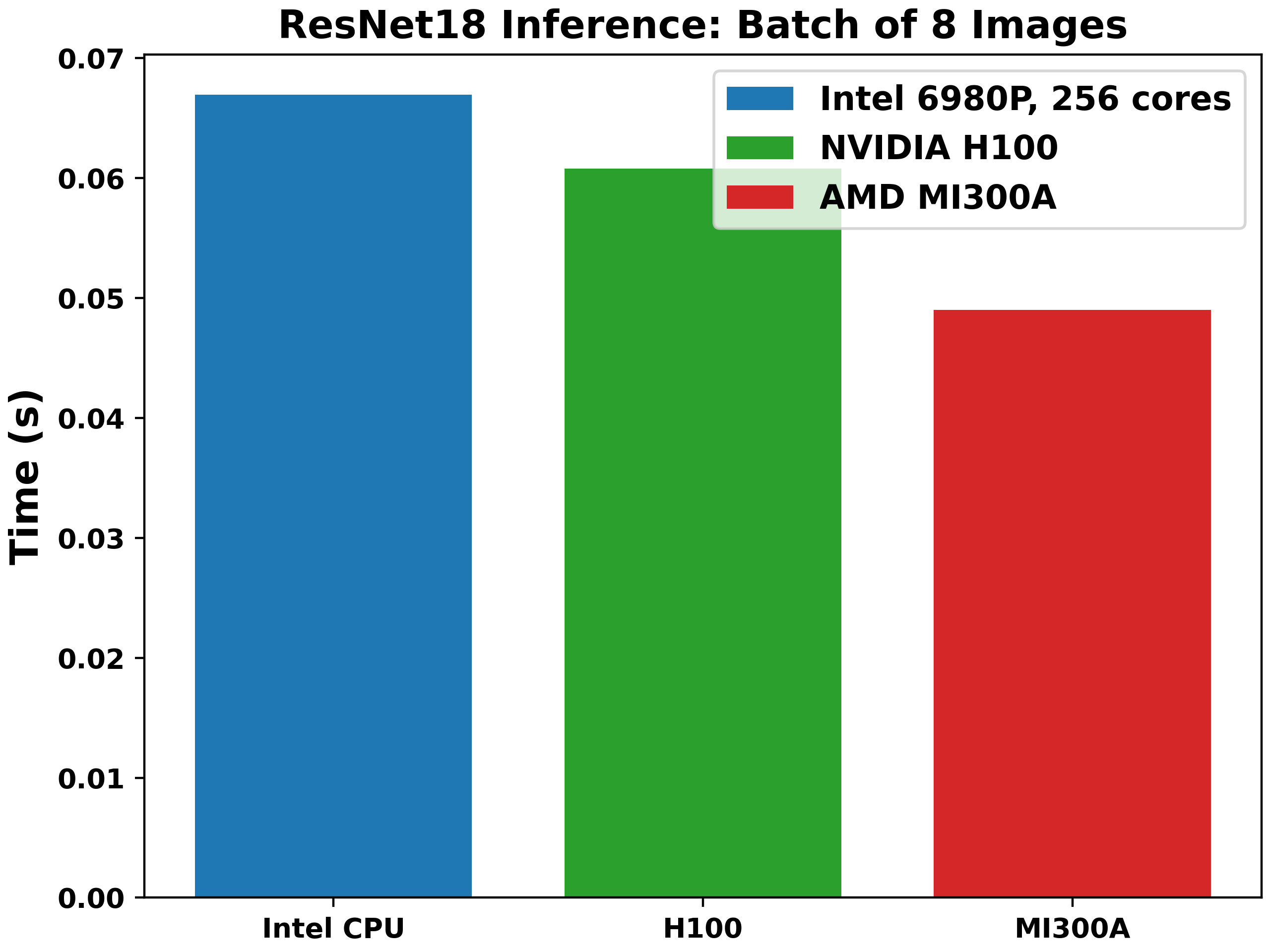}
    \caption{ResNet18 inference.}
    \label{fig:resnet}
    \end{subfigure}
    \caption{CSE and AI/ML Integration performance: (a) MALA inference performance on a batch of 8748 atoms; (b) ResNet18 inference performance on a batch of 8 images. (lower is better)}
\end{figure}

\paragraph{MALA}
We describe the MALA benchmark before presenting the results. Computing the electronic structure of materials is critical for understanding material  behavior at different temperatures and pressure, designing new materials, drug discovery and number of other applications. The most widely used method for electronic structure calculation is density functional theory (DFT). However, DFT scales O($N^3$) where $N$ is the number of atoms. This limits the use of DFT to small systems. SciML methods, especially physics-informed deep neural networks in MALA (MAterial Learning Algorithms) \cite{fiedler2023predicting,cangi2024materials}, enable building a surrogate for the DFT calculations. For a DFT calculation using $N$ atoms where $N=1000$, MALA uses a $n_k \times n_k \times n_k$ grid where $n_k=256$ for evaluating local density of states at each grid point. These local quantities of interest are then used to compute global quantities such as band energy. The key message is there are more than 16 million inference (for $n_k=256$) that needs to be done in parallel. Lowering the physics-informed DNN surrogate to Kokkos C++ code allows the inference to be run from other codes like LAMMPS. LAPIS is able to lower the DNN in MALA and compute the band energy accurately. This provides a portable, high productivity way for developing an ML surrogate in python and deploying the surrogate within Kokkos C++ codes. Figure \ref{fig:mala} shows the results of automatically lowered MALA DNN running on three different architectures. Notice than MALA inference used 8748 atoms for about 0.5 ms. \textit{The DFT calculation of similar scale is beyond the reach of most large cluster installations}. LAPIS enables MALA  on three different architectures by automatically lowering the ML models.

\paragraph{ResNet18}
We used a pre-trained image classification CNN model from PyTorch, ResNet18, as the target for testing LAPIS on a traditional ML use case. Section \ref{sec:aiml_demo} described the lowering process for the ResNet18 model.
We were successful in creating a fully automated pipeline that generates working Kokkos C++ source code from the PyTorch model. This code can successfully perform ResNet18 inference on real images when integrated into a standalone C++ program. As in MALA, LAPIS based code generation also allows transforming Python sources to multiple hardware targets such as CPUs and GPUs.  We demonstrate transformations to HIP, OpenMP and CUDA backends of generated Kokkos code. The generated code was evaluated on the three architectures of interest. Figure \ref{fig:resnet} shows the portable execution of ResNet18 on Intel CPU, H100 and MI300A for a batch of eight images.  This performance includes the lowering of all the ResNet18 layers such as 2D convolutions, batch normalization, ReLU, max-pooling etc. The Kokkos inspired memory references are critical for performance of such an architecture so we avoid memory copies between host and device for every one of the layers.

\subsection{Dense linear algebra}

\begin{table}[]
\begin{tabular}{llll}
Architecture & LAPIS (ms) & KokkosKernels (ms) \\ \hline
Granite Rapids & 1.83 & 1.86  \\
H100 & 0.359 & 0.358  \\
MI300A & 0.264 & 0.263 \\
\end{tabular}
\caption{Time in ms to multiply two 4096x4096 FP32 matrices (SGEMM), using LAPIS (with vendor library calls enabled) and with Kokkos Kernels. No measurable overhead is introduced by LAPIS.}
\label{tab:gemm}
\end{table}

\begin{figure}
    \centering
    \includegraphics[width=0.75\linewidth]{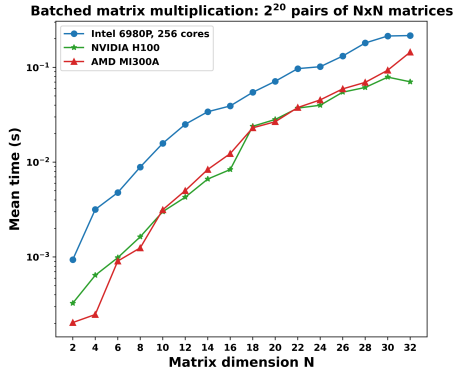}
    \caption{Batched GEMM performance across the 3 backends (lower is better). Times were averaged over 100 trials.}
    \label{fig:batched_gemm}
\end{figure}

Dense linear algebra is fundamental to many use cases in CSE and AI/ML. The compute bound kernels in this space are typically hand-optimized for achieving near peak performance of an architecture. LAPIS can lower the dense matrix multiplies represented typically with a single line of code \texttt{torch.matmul(a, b)} where \texttt{a} and \texttt{b} are dense tensors to Kokkos-based hierarchical parallel nested loops. It is possible to intercept these operations that we know are hand-optimized and call libraries such as cuBLAS and rocBLAS. However, we use these two dense linear algebra operations to show the LAPIS functionality and portability for general dense kernels. 
GEMM or General-Matrix-Matrix multiplication is a fundamental Basic Linear Algebra Subroutine \cite{blackford2002updated} kernel to build other fundamental operations such as those in LAPACK \cite{anderson1999lapack}. Numerical vendor libraries include highly optimized, platform-specific implementations of GEMM, so the Kokkos Kernels  library \cite{rajamanickam2021kokkos} provides portable wrappers for several common libraries. While LAPIS's lowering pipeline can generate a portable GEMM in pure Kokkos, we also support generating calls to Kokkos Kernels instead (which in turn calls the appropriate library). Table \ref{tab:gemm} shows that when a LAPIS-generated program calls MKL, cuBLAS or rocBLAS GEMM through Kokkos Kernels, no noticeable overhead is introduced. This allows LAPIS-generated C++ code to maintain portability while achieving the same performance as platform-specific libraries on common operations. Currently, LAPIS supports wrapper calls to GEMM and GEMV (matrix-vector multiplication) with FP16, FP32 and FP64 data types. This approach could be applied to all kernel interfaces provided by Kokkos Kernels, however.

A more recent addition to dense linear algebra standards is batched linear algebra kernels \cite{abdelfattah2021set} that operate on large number of small/medium size dense matrices. While the original motivation was from AI/ML codes, batched linear algebra has become tool for CSE codes as well. We use the batched GEMM as a demonstrator for LAPIS' capabilities to lower batched kernels. Batched kernels specifically require ability to handle hierarchical parallelism and multi-dimensional arrays with the correct batch dimension on different architectures. For example, it is critical to vectorize on the batch dimension as the matrix dimensions are quite small. LAPIS is able to lower the batched GEMM in a portable manner as demonstrated in Figure \ref{fig:batched_gemm}.

\section{Conclusion and Future Work}
\label{sec:concl}


We demonstrate that LAPIS can generate performance, portable code from high-productivity language like python. As a result, LAPIS, retains the performance portability advantages of frameworks like Kokkos and
productivity of PyTorch. Furthermore, we demonstrate that a compiler framework, LAPIS, using MLIR can automate the process of integration between CSE and AI/ML codes.  LAPIS allows domain scientists to exchange data seamlessly between simulation codes in C++ and machine learning models from Python. Our compiler can take a general PyTorch model and generate parallel, portable C++ source code that uses the Kokkos programming model, making it trivial to integrate into existing Kokkos-based applications that can run on different CPUs and GPUs.
LAPIS, with its perforrmance portable additions
adds significant new capabilities to MLIR. All of these features can impact multiple domain specific compilers that are built using MLIR.
In the future, we plan to explore other front-ends beyond PyTorch and torch-mlir such as TensorFlow \cite{tensorflow}, JAX \cite{jax} and IREE \cite{IREE}. We also plan to expand the number of kernels using LAPIS to other ML models and linear algebra kernels.




\section*{Acknowledgments.}
We would like to thank Kim Liegeois for naming LAPIS and for his help in the early stages of its development. We also thank Miheer Vaidya for his help in identifying and fixing issues in LAPIS. Finally, we thank P. Sadayappan and Atanas Rountev for sharing their advice and compiler expertise. 

This work was supported by the Laboratory Directed Research and Development program at Sandia
National Laboratories, a multimission laboratory managed and operated by National Technology and
Engineering Solutions of Sandia LLC, a wholly owned subsidiary of Honeywell International Inc. for the U.S.
Department of Energy’s National Nuclear Security Administration under contract DE-NA0003525. The work was also supported by the S4PST project, under the U.S. Department of Energy's Office of Science Advanced Scientific Computing Research (ASCR) program.

\bibliographystyle{siamplain}
\bibliography{main}

\appendix
\section{Appendix A: Supplementary Material}
\subsection{LAPIS User Interface}
We describe the ways to use LAPIS in this subsection.
The LAPIS software provides two interfaces into the pipeline. One is a pair of command-line utilities \texttt{lapis-opt} (for lowering linalg-on-tensors IR to the Kokkos dialect) and \texttt{lapis-translate} (for running the Kokkos emitter). These two utilities support the same basic interface as the built-in tools \texttt{mlir-opt} and \texttt{mlir-translate}, including the ability to read input from stdin and write output to stdout. Given some valid input IR, the pipeline can be executed as:
\begin{verbatim}
cat input.mlir |\
lapis-opt --sparse-compiler-kokkos |\
lapis-translate -o output.cpp
\end{verbatim}

In addition to the command-line interface, LAPIS provides a Python package with a \texttt{KokkosBackend} class to drive the lowering pipeline and the compilation of the generated C++ into native code. A \texttt{ctypes}-based Python wrapper is also generated to allow this native code to be called directly from Python. This is functionally a drop-in replacement for the RefBackend in torch-mlir \cite{torchmlir}, as long as Kokkos is installed ahead of time (the environment variable \texttt{KOKKOS\_ROOT} is used to locate the desired installation). This fully end-to-end mode is useful for testing LAPIS and the correctness of the generated code in one script.

Given a Torch module \verb|m| and a set of input tensors (here a, b, c), the following Python code will perform the entire end-to-end compilation process on the module's \texttt{forward} method, and then execute it. When providing specific tensor objects, the tensor parameters in the generated code will use matching compile-time constant shapes.

\begin{verbatim}
    from lapis import KokkosBackend
    ...
    as_mlir = \
      torchscript.compile(m, (a, b, c), \
      output_type='linalg-on-tensors')
    m_compiled =\
      KokkosBackend.KokkosBackend(). \
      compile(as_mlir)
    results = m_compiled.forward(a, b, c)
\end{verbatim}

To generate code with dynamic shapes instead, torch-mlir provides a tensor placeholder object where a dimension of -1 indicates that it is dynamic. A single tensor can have mixed static and dynamic dimensions. This can be passed to \texttt{torchscript.compile} instead of an actual tensor as follows.
\begin{verbatim}
    from torch_mlir.compiler_utils \
      import TensorPlaceholder
    ...
    a = TensorPlaceholder(\
      [-1, -1, 100], torch.float)
\end{verbatim}


\subsection{System and Software Configuration}
\label{sec:sysconfig}

The configurations and system information used for these three high performance computing relevant backends in our experiments are as follows:

\begin{itemize}
    \item \textbf{OpenMP:} Blake testbed with dual Intel 6980P (Granite Rapids) CPUs. The compiler was Intel icpx 2024.1.0. We used OMP\_NUM\_THREADS=128, OMP\_PLACES=cores, OMP\_PROC\_BIND=close, to utilize all 128 cores on one socket without hyperthreading, since using both sockets or hyperthreading resulted in noisy timing.
    \item \textbf{CUDA:} Hops cluster with four NVIDIA H100 GPUs on each node (one was used during experiments). The compiler/runtime was CUDA 12.4.0.
    \item \textbf{HIP:} RZAdams cluster with four AMD MI300A APUs on each node (one APU was used during experiments). The compiler/runtime was \texttt{hipcc} and ROCm 6.4.2.
\end{itemize}


Since portability and automation are key features of LAPIS, the LAPIS-generated C++ files for each of the following tests were generated once and compiled on all three target systems with no modifications. A C++ \texttt{main()} function was manually written for each test to read input data from a file and call the kernels in a timed loop. Unless otherwise specified, all tests use one untimed warm-up call and then average the running time over 1000 repetitions.

\end{document}